\documentclass[aps,prl,floatfix,twocolumn,superscriptaddress]{revtex4-2}

\usepackage{bm}
\newcommand{\rr}{\mathbf{r}}
\usepackage{graphicx,
            caption,
            subcaption,
            amsmath,
            braket,
            units,
            ragged2e,
            xcolor,
            xspace,
            nicefrac,
            lipsum,
            nameref,
            textcomp,
            urwchancal,
            upgreek,
            isotope,
            hyperref,
            cleveref,
            siunitx,
            placeins, 
            scrextend, 
            url}

\DeclareCaptionLabelSeparator{dot}{. }
\makeatletter
\def\justified{
	\let\\\@normalcr
	\@rightskip\z@skip \rightskip\@rightskip
	\leftskip\z@skip
	\parindent 0em\relax
	\setlength{\parfillskip}{0pt plus 1fil}}
\DeclareCaptionJustification{justified}{\justified}
\hypersetup{colorlinks=true, linkcolor=blue,citecolor=blue,urlcolor=blue}
\def\unit #1 #2 {\SI{#1}{#2}\xspace}
\def\range #1 #2 #3 {\SIrange{#1}{#2}{#3}\xspace}
\DeclareSIUnit\gauss{G}

\newcommand{\myref}[2][]{Fig.~\hyperref[#2]{\ref*{#2}#1}}
\newcommand{\Myref}[2][]{Figure~\hyperref[#2]{\ref*{#2}#1}}
\newcommand{\Mytabref}[2][]{Table~\hyperref[#2]{\ref*{#2}#1}}

\begin{document}
\linespread{1.5}
\title{Two-dimensional supersolidity in a dipolar quantum gas}

\date{\today}


 \author{Matthew A. Norcia}
 \thanks{M.~A.~N.~and C.~P.~contributed equally to this work.}
 \affiliation{
     Institut f\"{u}r Quantenoptik und Quanteninformation, \"Osterreichische Akademie der Wissenschaften, Innsbruck, Austria
 }

 \author{Claudia Politi}
  \thanks{M.~A.~N.~and C.~P.~contributed equally to this work.}
 \affiliation{
     Institut f\"{u}r Quantenoptik und Quanteninformation, \"Osterreichische Akademie der Wissenschaften, Innsbruck, Austria
 }
 \affiliation{
     Institut f\"{u}r Experimentalphysik, Universit\"{a}t Innsbruck, Austria
 }

 \author{Lauritz Klaus}
 \affiliation{
     Institut f\"{u}r Quantenoptik und Quanteninformation, \"Osterreichische Akademie der Wissenschaften, Innsbruck, Austria
 }
 \affiliation{
     Institut f\"{u}r Experimentalphysik, Universit\"{a}t Innsbruck, Austria
 }

 \author{Elena Poli}
 \affiliation{
     Institut f\"{u}r Experimentalphysik, Universit\"{a}t Innsbruck, Austria
 }

\author{Maximilian Sohmen}
\affiliation{
    Institut f\"{u}r Quantenoptik und Quanteninformation, \"Osterreichische Akademie der Wissenschaften, Innsbruck, Austria
}
\affiliation{
    Institut f\"{u}r Experimentalphysik, Universit\"{a}t Innsbruck, Austria
}

 \author{Manfred J. Mark}
 \affiliation{
     Institut f\"{u}r Quantenoptik und Quanteninformation, \"Osterreichische Akademie der Wissenschaften, Innsbruck, Austria
 }
 \affiliation{
     Institut f\"{u}r Experimentalphysik, Universit\"{a}t Innsbruck, Austria
 }

 \author{Russell Bisset}
 \affiliation{
     Institut f\"{u}r Experimentalphysik, Universit\"{a}t Innsbruck, Austria
 }

 \author{Luis Santos}
 \affiliation{
     Institut f\"{u}r Theoretische Physik, Leibniz Universit\"{a}t Hannover, Germany
 }

 \author{Francesca Ferlaino}
 \thanks{Correspondence should be addressed to \mbox{\url{Francesca.Ferlaino@uibk.ac.at}}}
 \affiliation{
     Institut f\"{u}r Quantenoptik und Quanteninformation, \"Osterreichische Akademie der Wissenschaften, Innsbruck, Austria
 }
 \affiliation{
     Institut f\"{u}r Experimentalphysik, Universit\"{a}t Innsbruck, Austria
 }


\begin{abstract}

Supersolidity --- a quantum-mechanical phenomenon characterized by the presence of both superfluidity and crystalline order --- was initially envisioned in the context of bulk solid helium, as a possible answer to the question of whether a solid could have superfluid properties \cite{Gross:1957, Gross:1958, Andreev:1969, Chester:1970, Leggett:1970}.    While supersolidity has not been observed in solid helium (despite much effort)\cite{chan2013overview}, ultracold atomic gases have provided a fundamentally new approach, recently enabling the observation and study of supersolids with dipolar atoms \cite{Lu:2015, Biallie:2018, Roccuzzo:2019, Boninsegni:2012, Tanzi:2019, Bottcher2019, Chomaz:2019, guo2019low, Natale2019, tanzi2019supersolid}.  However, unlike the proposed phenomena in helium, these gaseous systems have so far only shown supersolidity along a single direction.  By crossing a structural phase transition similar to those occurring in ionic chains \cite{birkl1992multiple, Raizen1992, Fishman:2008, Shimshoni2011}, quantum wires \cite{Hew:2009, Mehta:2013qwires}, and theoretically in chains of individual dipolar particles \cite{Astrakharchik:2008, Ruhman:2012}, we demonstrate the extension of supersolid properties into two dimensions, providing an important step closer to the bulk situation envisioned in helium.  This opens the possibility of studying rich excitation properties \cite{Santos:2003, Ronen:2007, Wilson:2008, Bisset:2013}, including vortex formation \cite{Gallem:2020, Roccuzzo:2020, ancilotto2020vortex}, as well as ground-state phases with varied geometrical structure \cite{Lu:2015, Zhang:2019} in a highly flexible and controllable system.

\end{abstract}

\maketitle

\begin{figure}[ht]
    \centering
	\includegraphics[width=3.38in]{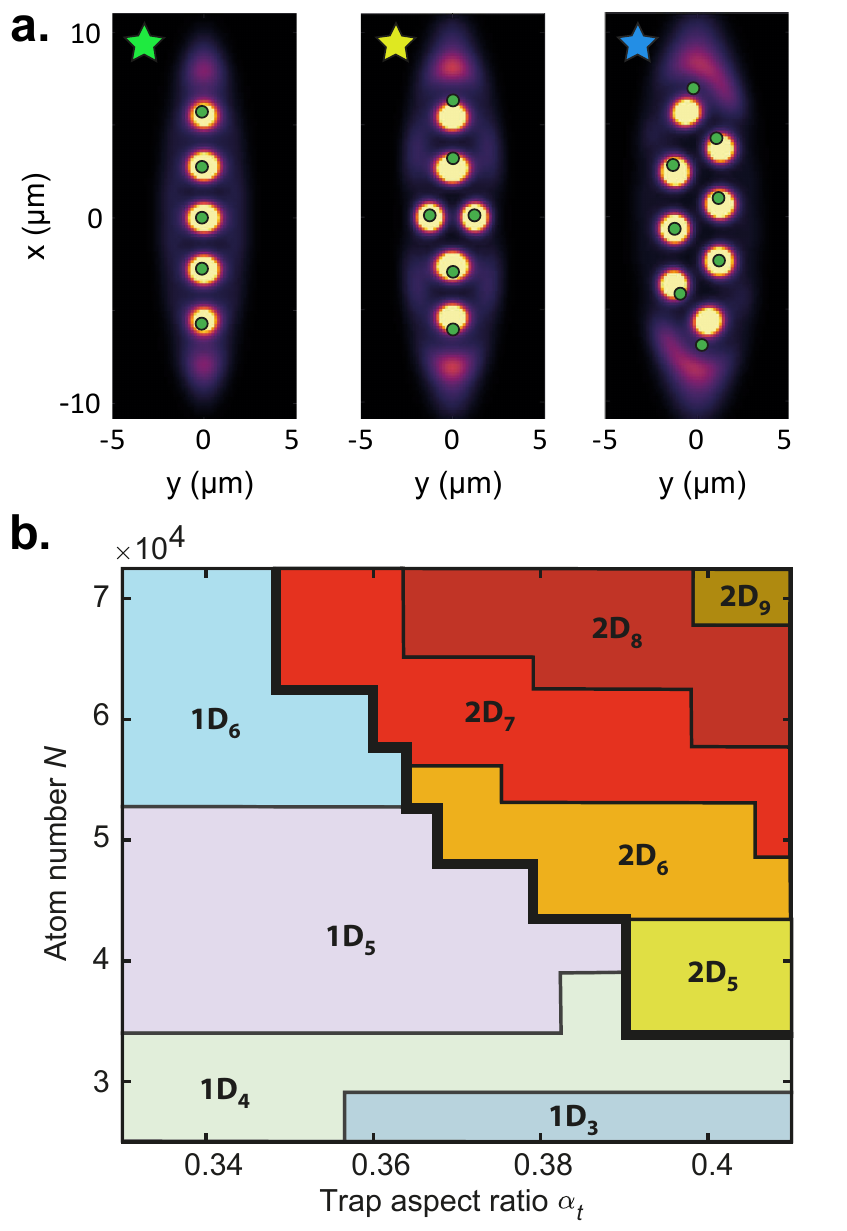}
\caption {
	{\bf Calculated phases of dipolar droplet array.}
	\textbf{a.} In-trap ground-state density profiles calculated using eGPE for atom numbers $N \in [3.3, 4.4, 5.8] \times 10^4$ in the droplets and trap aspect ratios $\alpha_t = f_x/f_y \in [0.33, 0.35, 0.39]$ (left to right).  The scattering length $a=88\,a_0$, where $a_0$ is the Bohr radius.  Green dots depict the droplet positions obtained from the variational model, assuming the same $N$ and droplet number $N_D$ as the eGPE.  Stars connect to experimentally observed density profiles in Fig.\,\ref{fig:2}b.  
	 \textbf{b.}  Phase diagram, obtained from our variational model, as a function of $N$ and $\alpha_t$ for $f_x=33$\,Hz, $f_z=167$\,Hz.   Linear~(two-dimensional) phases with $N_D$ droplets are labelled as 1D$_{N_D}$~(2D$_{N_D}$).} 
	 \label{fig:1} 
\end{figure}

Ultracold atoms have recently offered a fundamentally new direction for the creation of supersolids  --- rather than looking for superfluid properties in a solid system like $^4$He, ultracold atoms allow one to induce a crystalline structure in a gaseous superfluid, a system which provides far greater opportunity for control and observation.  This new perspective has enabled supersolid properties to be observed in systems with spin-orbit coupling \cite{Li:2017} or long-range cavity-mediated interactions \cite{Leonard:2017}, though in these cases the crystalline structure is externally imposed, yielding an incompressible state.  In contrast, 
dipolar quantum gases of highly magnetic atoms can spontaneously form crystalline structure due to intrinsic interactions \cite{Tanzi:2019, Bottcher2019, Chomaz:2019}, allowing for a supersolid with both crystalline and superfluid excitations \cite{guo2019low, Natale2019, tanzi2019supersolid}.  In these demonstrations, supersolid properties have only been observed along a single dimension, as a linear chain of phase-coherent ``droplets", i.e.~regions of high density connected by low-density bridges of condensed atoms, confined within an elongated optical trap.    

The extension of supersolidity into two dimensions is a key step towards creating an ultracold gas supersolid that is closer to the states envisioned in solid helium.  
Compared to previous studies of incoherent two-dimensional dipolar droplet crystals \cite{kadau2016observing, Biallie:2018}, we work with both a substantially higher atom number $N$ and relatively strong repulsive contact interactions between atoms.  This leads to the formation of large numbers of loosely bound droplets, enabling us to establish phase coherence in two dimensions.  
In our system, the repulsive dipolar interactions between droplets facilitate a structural transition from a linear to a two-dimensional array, analogous to the Coulomb-interaction-mediated structural phase transitions observed with ions \cite{birkl1992multiple, Raizen1992,Fishman:2008,  Shimshoni2011}. Unlike ions however, our droplets are compressible and result from the spontaneous formation of a density wave, allowing for dynamical variation in both droplet number and size.  Further, the exchange of particles between droplets enables the spontaneous synchronization of the internal phase of each droplet across the system, and the associated superfluid excitations \cite{guo2019low, Natale2019, tanzi2019supersolid}.  

\begin{figure*}[ht]
    \centering
	\includegraphics[width=\textwidth]{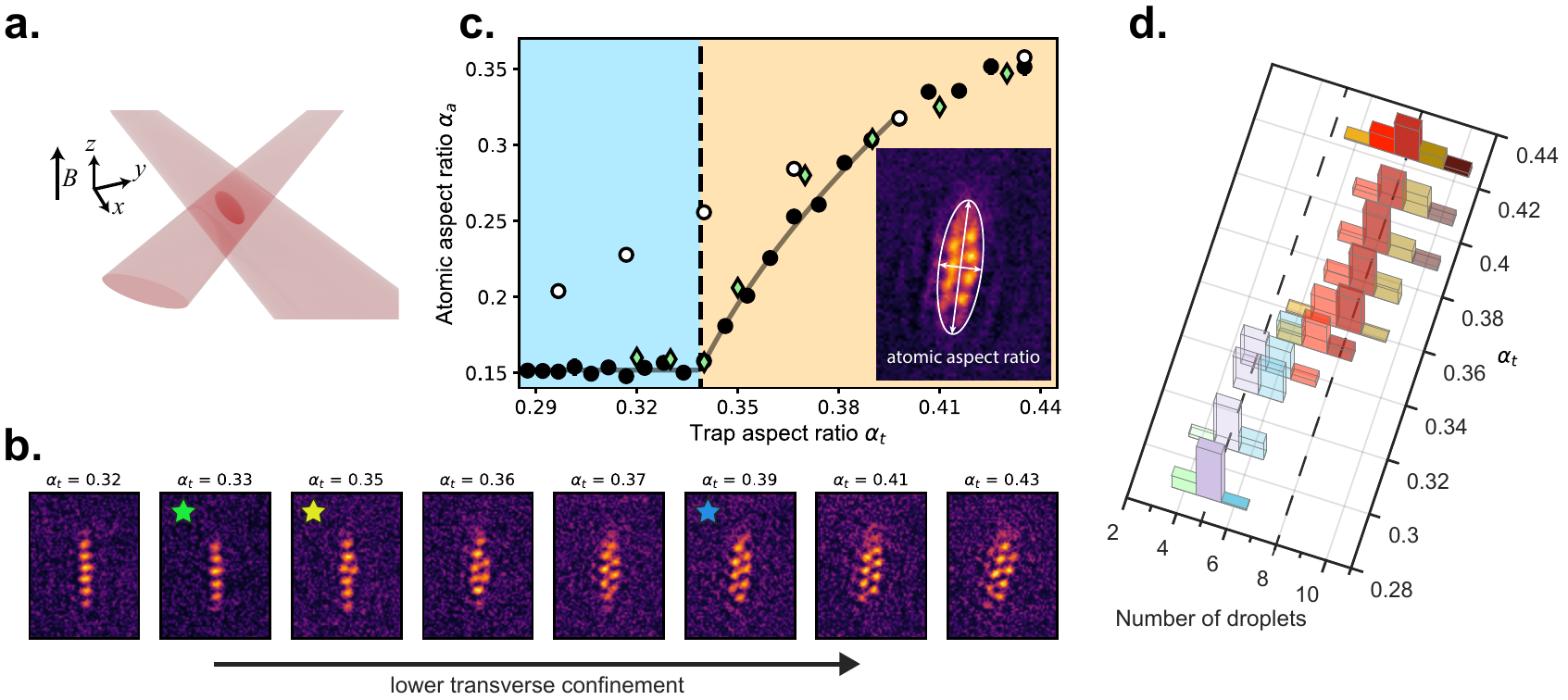}
	\caption {
	{\bf Linear to zig-zag transition in an anisotropic trap.}
	\textbf{a.}  We confine and condense dipolar $^{164}$Dy atoms within an anisotropic optical dipole trap (ODT) formed by the intersection of two laser beams.  By tuning the aspect ratio of the trap in the $x$-$y$ plane ($\alpha_t$), perpendicular to an applied magnetic field $B$, we induce a transition between linear and zig-zag configurations of droplets.  
	\textbf{b.} Single-trial images of the in-trap density profile of atoms at different $\alpha_t$, showing structural transition from linear to zig-zag states, as well as an increase in droplet number for higher $\alpha_t$.  Stars indicate values $\alpha_t$ and $N$ corresponding to the eGPE calculations of Fig.~\ref{fig:1}a.  
	\textbf{c.}  Atomic aspect ratio $\alpha_a$ versus trap aspect ratio $\alpha_t$.  $\alpha_a$ is the ratio of minor to major axes of a two-dimensional Gaussian fit to the imaged in-trap density profile (inset).  For the supersolid droplet array (black markers) we see an abrupt change in $\alpha_a$ at the critical trap aspect ratio $\alpha_t^*$, extracted from the fit (gray line, see methods).  The shape of the transition agrees well with eGPE prediction (green diamonds, see methods).  For an unmodulated condensate (white markers), no abrupt change is evident.  
	\textbf{d.} Distribution of droplet number versus $\alpha_t$, showing a distinct increase in droplet number at the transition of linear to zig-zag configurations.  
	}
	 \label{fig:2} 
\end{figure*}


Dipolar quantum gases exhibit a rich set of ground- and excited-state phenomena due to the competition between many energetic contributions.  These include mean-field interactions of both contact and dipolar nature, quantum fluctuations, and external confinement, parameterized by potentially anisotropic trapping frequencies $f_{x, y, z}$.  
Such systems can be described with great accuracy by using an extended Gross--Pitaevskii equation~(eGPE) \cite{FerrierBarbut:2016, Chomaz:2016, Wachtler2016, Bisset:2016}.  Even a fine variation of the strength of these energetic contributions can lead to dramatic qualitative changes in the state of the system, for example enabling a transition from a uniform condensate to a supersolid, or in our present case, from a linear supersolid to a two-dimensional one.  

Fig.\,\ref{fig:1}a shows ground-state density profiles calculated across this transition using the eGPE at zero temperature.  These profiles feature arrays of high-density droplets, immersed in a low-density coherent ``halo" that establishes phase-coherence across the system.  As the trap becomes more round, the initially linear chain of droplets acquires greater transverse structure, eventually forming a zig-zag state consisting of two offset linear arrays.  

Although the eGPE has remarkable predictive power, full simulations in three dimensions are numerically intensive, making a global survey of the array properties as a function of our experimental parameters difficult.
To overcome this limitation, we employ a variational ansatz that captures the key behavior of the system, and allows us to disentangle the competing energetic contributions.  
In this approach, we describe an array of $N_D$ droplets by the wavefunction 
$\psi(\rr)=\sum_{j=1}^{N_D} \psi_j(\rr)$, where the  $j$--th droplet is assumed to be of the form: 
$\psi_j(\rr) \propto \sqrt{N_j}\exp \left (-\frac{1}{2}\left( \frac{|\bm{\rho}-\bm{\rho}_j|}{\sigma_{\rho,j}}\right)^{r_{\rho,j}} \right ) \exp \left (-\frac{1}{2}\left( \frac{|z-z_j|}{\sigma_{z,j}} \right)^{r_{z,j}} \right )$,
interpolating between a Gaussian and a flat-top profile characteristic of quantum droplets~\cite{Lavoine2020}. 
For a given total number of atoms $N$ and droplet number $N_D$, energy minimization provides the atom number $N_j$ in each droplet, as
well as their widths $\sigma_{\rho(z),j}$, exponents $r_{\rho(z),j}$, and positions $\bm{\rho}_j=(x_j,y_j)$. Repeating this energy minimization as a function of $N_D$ gives the optimal number of droplets. 
This model provides a good qualitative description of the overall phase diagram~(Fig.~\ref{fig:1}b), revealing that the interplay between intra-droplet physics 
and inter-droplet interaction results in a rich landscape of structural transitions as a function of the atom number and the trap aspect ratio $\alpha_t = f_x/f_y$. 

Several trends are immediately visible from the phase diagram.  Larger $N$ and higher $\alpha_t$ generally produce states with larger numbers of droplets.  Further, as with ions, a large number of droplets favors a 2D configuration, while tighter transverse confinement (small $\alpha_t$) favors 1D \cite{birkl1992multiple, Raizen1992, Fishman:2008, Shimshoni2011}.  A transition from 1D to 2D is thus expected when moving towards larger $N$ or to higher $\alpha_t$.  
In stark contrast to the case of ions, the number of droplets typically increases across the 1D to 2D transition, implying a first-order nature, while only narrow regions in the phase diagram may allow for a 1D-to-2D transition at constant droplet number.  

The variational results are in excellent agreement with our eGPE numerics, in terms of predicting the qualitative structure of droplet array patterns, as shown in Fig.~\ref{fig:1}a.
Slight discrepancies exist between the two theories regarding the predicted droplet positions and the location of the 1D-to-2D transition.  This is likely because of the presence of the halo in the eGPE simulation (and presumably in the experiment), visible in Fig.\,\ref{fig:1}a, which is not accounted for in the variational model.  This halo appears to accumulate at the ends of the trap, pushing the droplets toward the trap center and likely increasing the effective trap aspect ratio experienced by the droplets.  

To explore the 1D to 2D transition experimentally, we use a condensate of highly magnetic $^{164}$Dy atoms confined within an anisotropic optical dipole trap with independently tunable trap frequencies $f_{x, y, z}$.  The trap, shown in Fig.\,\ref{fig:2}a, is shaped like a surf-board with the tight axis along gravity and along a uniform magnetic field that orients the atomic dipoles and allows tuning of the contact interaction strength.  Typically, we perform evaporation directly into our state of interest at our desired final interaction strength, as demonstrated in Refs.\,\cite{Chomaz:2019, sohmen2021birth}.  A combination of in-trap and time-of-flight (TOF) imaging provides us with complementary probes of the density profile of our atomic states, and the phase coherence across the system.


\begin{figure*}[ht]
    \centering
	\includegraphics[width=\textwidth]{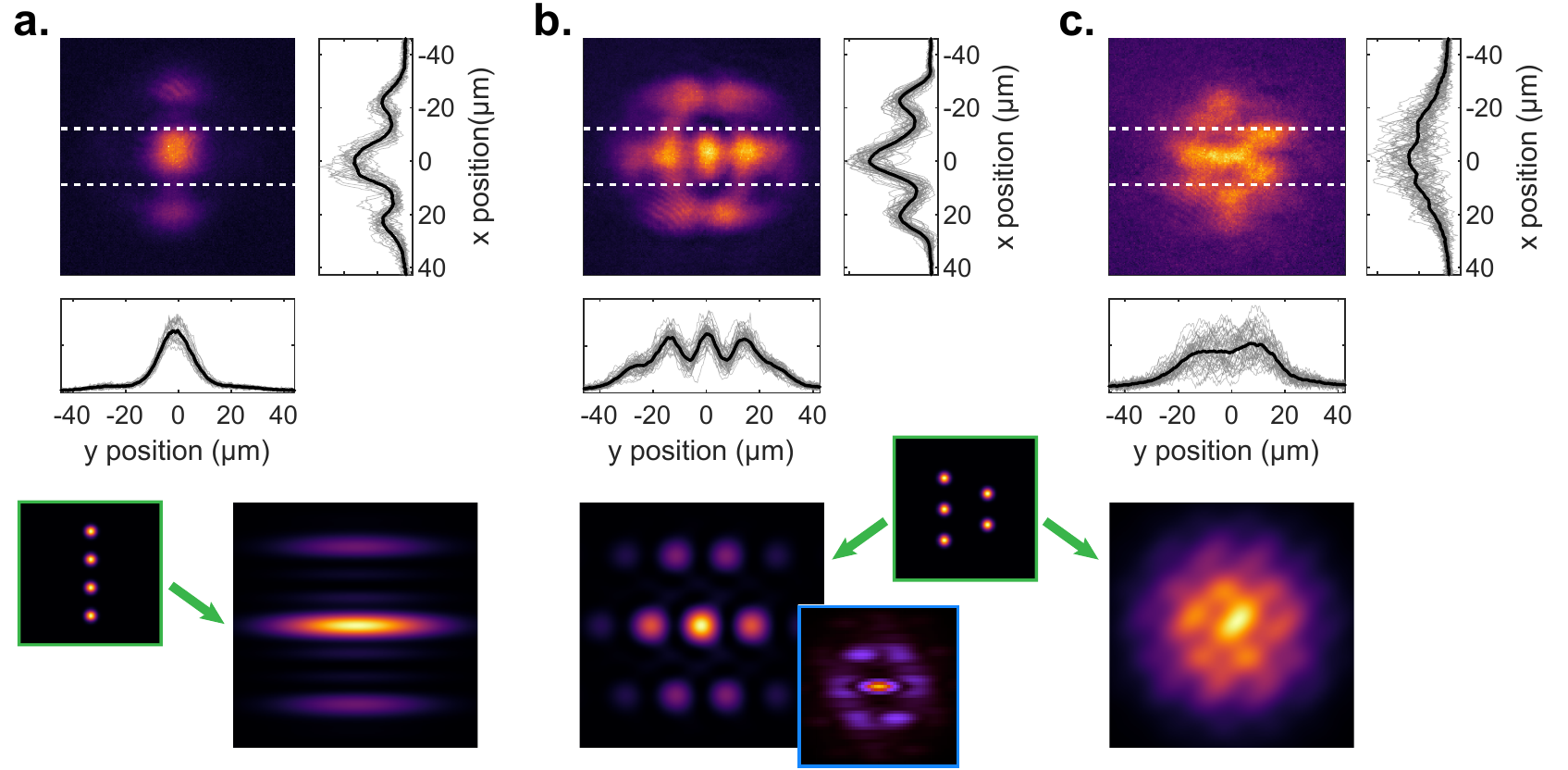}
	\caption {
	{\bf Coherence in linear and zig-zag states.}  Upper panels show averaged images of experimental TOF interference patterns, along with projections along horizontal and vertical directions of average (solid black lines) and individual images (gray lines).  The vertical projection is calculated between the dashed lines.  Lower panels show interference patterns calculated for the pictured in-trap droplet configurations (green outlines).  
	\textbf{a.}  Linear chain of phase-coherent droplets, showing uniaxial modulation persisting in averaged image (26 trials).  
	\textbf{b.}  Zig-zag configuration of phase-coherent droplets, showing modulation along two directions that persists in averaged image (51 trials), and hexagonal structure.  The spacing of rows in the simulation was adjusted to approximate the observed aspect ratio of TOF image.  The image outlined in blue shows the average momentum distribution calculated from a series of 20 variational calculations converging to slightly different droplet configurations, showing the tendency of such fluctuations to broaden features in the interference pattern while maintaining the underlying structure.  
	\textbf{c.}  Zig-zag configuration of phase-incoherent droplets.  Modulation remains in single images, as evidenced by the spread of gray traces in projection, but washes out in average (43 trials).  
	}
	 \label{fig:4} 
\end{figure*}

We begin by studying the transition from one to two dimensions by changing the strength of transverse confinement provided by the trap.  Our optical setup allows us to tune $f_y$ from roughly 75 to 120~Hz, while leaving $f_x$, $f_z$ nearly constant at 33(2), 167(1)~Hz, and thus to vary the trap aspect ratio $\alpha_t$ in the plane perpendicular to the applied magnetic field and our imaging axis.  For small $\alpha_t$, the atoms are tightly squeezed transversely, and form a linear-chain supersolid (as seen in in-trap images of Fig.\,\ref{fig:2}b).  As we increase $\alpha_t$ above a critical value $\alpha^*_t = 0.34(2)$, we observe a structural phase transition to a two-dimensional (2D) state with two side-by-side droplets in the center of the chain.  By further increasing $\alpha_t$, the 2D structure extends to two offset lines of droplets in a zig-zag configuration.  The observed patterns match well with the ground-state predictions from the eGPE calculations when we globally fix the scattering length to 88$a_0$.

We obtain higher atom numbers in the more oblate traps (higher $\alpha_t$), giving $N =  6.5(5) \times 10^4$ at $\alpha_t = 0.44$ and $N = 2.5(4) \times 10^4$ at $\alpha_t = 0.28$.  This further facilitates the crossing of the 1D to 2D transition, by favoring states with larger numbers of droplets in the broader traps.  In the zig-zag regime, two-dimensional modulation is clearly visible for durations beyond one second.  Further, the droplet configuration patterns are fairly repeatable, with clear structure visible in averaged images as shown in the inset of Fig.\,\ref{fig:2}c, which is an average of 23 trials taken over roughly two hours.  

The transition from 1D to 2D is immediately visible when plotting the atomic aspect ratio $\alpha_a$ versus $\alpha_t$, as shown in Fig.\,\ref{fig:2}c.
We find that $\alpha_a$ undergoes a rapid change at $\alpha^*_t$, as the single linear chain develops two-dimensional structure.  For comparison, we plot $\alpha_a$ measured for an unmodulated BEC, formed at a different magnetic field, which does not feature the sharp kink present for the supersolid state.

In Fig.\,\ref{fig:2}d, we show the number of droplets present for different $\alpha_t$.  In the 1D regime, we typically see between five and six droplets.  This number abruptly jumps up by approximately one droplet for 2D states near the transition point, and then increases up to an average value of eight droplets as $\alpha_t$ is further increased.  The change in droplet number indicates that the transition that we observe is not of simple structural nature, but is also accompanied by a reconfiguration of atoms within the droplets, as expected from theory (see Fig.\,\ref{fig:1}).


The measurements of in-trap density presented above inform us about the structural nature of the transition, but not about phase coherence, which is the key distinguishing feature between an incoherent droplet crystal and a supersolid.  
Previous observations of 2D droplet arrays \cite{kadau2016observing} were performed in traps where the ground state is a single droplet \cite{Biallie:2018}, and the observed droplet crystal was likely a metastable state lacking inter-droplet phase coherence.  
In contrast, we expect from our theoretical calculations that the 2D array is the ground state of our surfboard-shaped trap (for $\alpha_t > \alpha^*_t$),  facilitating the formation of a phase-coherent, and therefore supersolid state for our experimental parameters.

We experimentally demonstrate the supersolid nature of our 2D modulated state using a matter-wave interference measurement, as previously used in linear supersolid chains \cite{Tanzi:2019, Bottcher2019, Chomaz:2019}, (Fig.\,\ref{fig:4}a).  In this measurement, an array of uniformly spaced droplets creates an interference pattern with spatial period proportional to the inverse of the in-trap droplet spacing.  The relative internal phase of the droplets determines both the contrast and spatial phase of the interference pattern \cite{hadzibabic2004interference}.  When averaging over many interference patterns, obtained on separate runs of the experiment, clear periodic modulation persists for phase-coherent droplets, but averages out if the relative droplet phases vary between experimental trials.  Thus, the presence of periodic modulation in an average TOF image provides a clear signature of supersolidity in our system, as it indicates both periodic density modulation and phase coherence.  

Figure\,\ref{fig:4}a shows an example of such an averaged interference pattern for a linear chain.  Uniaxial modulation is clearly present along the direction of the chain, indicating a high degree of phase coherence.  For comparison, we also show the expected interference pattern calculated for a linear array of four droplets from free-expansion calculations, showing similar structure.  

For conditions where in-trap imaging shows a 2D zig-zag structure, the averaged interference pattern exhibits clear hexagonal symmetry (Fig.\,\ref{fig:4}b).  This is consistent with our expectation, and is indicative of the triangular structure of the underlying state.  
To confirm that the observed modulation is not present without phase coherence, we repeat the measurement of Fig.\,\ref{fig:4}b at a magnetic field corresponding to independent droplets, and also compute averaged interference pattern for a zig-zag state with the phases of the individual droplets randomized between simulated trials (Fig.\,\ref{fig:4}c).  In both cases, the averaged image does not show clear periodic modulation.

By exploiting the transition between linear and zig-zag states, we have accessed a regime where the supersolid properties of periodic density modulation and phase coherence exist along two separate dimensions.  
Future work will focus on further understanding the spectrum of collective excitations in the full two-dimensional system \cite{Ronen:2007, Wilson:2008, Bisset:2013, schmidt2021roton}, where both the crystalline structure and the exchange of particles between droplets will play an important role.  
Further investigations may elucidate in more detail the nature of the phase transitions and expected configurations in a wider range of trap aspect ratios, as well as the role that defects play in the 2D system, either as phase-slips in the zig-zag patterns \cite{pyka2013topological, ulm2013observation}, or as vortices trapped between droplets of the array \cite{Gallem:2020, Roccuzzo:2020, ancilotto2020vortex}.

\begin{acknowledgments}
We thank the Innsbruck Erbium team and Blair Blakie for discussions. We acknowledge R.~M.~W.~van Bijnen for developing the code for our eGPE ground-state simulations.

\textbf{Author Contributions:} M.A.N, C.P., L.K., M.S., M.J.M and F.F. contributed experimental work.  E.P and R.B. performed eGPE calculations.  L.S. contributed variational model.  All authors contributed to interpretation of results and preparation of manuscript.  

\textbf{Funding:} The experimental team is financially supported through an ERC Consolidator Grant (RARE, No.\,681432), an NFRI grant (MIRARE, No.\,\"OAW0600) of the Austrian Academy of Science, the QuantERA grant MAQS by the Austrian Science Fund FWF No\,I4391-N. L.S and F.F. acknowledge the DFG/FWF via FOR 2247/PI2790. 
M.S.~acknowledges support by the Austrian Science Fund FWF within the DK-ALM (No.\,W1259-N27). 
L.S.~ thanks the funding by the Deutsche Forschungsgemeinschaft (DFG, German Research Foundation) under Germany’s Excellence Strategy – EXC-2123 QuantumFrontiers – 390837967. M.A.N.~has received funding as an ESQ Postdoctoral Fellow from the European Union’s Horizon 2020 research and innovation programme under the Marie Skłodowska‐Curie grant agreement No.~801110 and the Austrian Federal Ministry of Education, Science and Research (BMBWF). M.J.M.~acknowledges support through an ESQ Discovery Grant by the Austrian Academy of Sciences.  
We also acknowledge the Innsbruck Laser Core Facility, financed by the Austrian Federal Ministry of Science, Research and Economy. Part of the computational results presented have been achieved using the HPC infrastructure LEO of the University of Innsbruck.

\end{acknowledgments}

\bibliography{references}


\clearpage
\appendix
\renewcommand\thefigure{\thesection S\arabic{figure}}   
\setcounter{figure}{0}

\section{Methods}

\noindent \textbf{Experimental apparatus and protocols:} Our experimental apparatus has been described in detail in Ref.\,\cite{Trautmann2018}.  
Here, we evaporatively prepare up to $N=6.5(5) \times 10^4$ condensed $^{164}$Dy atoms in a crossed optical dipole trap formed at the intersection of two beams derived from the same $1064$\,nm laser, although detuned in frequency to avoid interference.  One beam (the static ODT) has an approximately $\SI{60}{\micro\meter}$ waist. The second (the scanning ODT) has an $\SI{18}{\micro\meter}$ waist, whose position can be rapidly scanned horizontally at 250\,kHz to create a variably anisotropic time-averaged potential.  By tuning the power in each beam, and the scanning range of the scanning ODT, we gain independent control of the trap frequencies in all three directions.  The two trapping beams propagate in a plane perpendicular to gravity, and cross at a $\SI{45}{\degree}$ angle, which leads to  the rotation of the zig-zag state at high $\alpha_t$ visible in Fig.\,\ref{fig:2}b.  

We apply a uniform magnetic field oriented along gravity and perpendicular to the intersecting dipole traps, with which we can tune the strength of contact interactions between atoms.  This allows us to create unmodulated Bose-Einstein condensates, supersolid states, or states consisting of independent droplets at fields of $B =$ 23.2\,G, 17.92\,G, and 17.78\,G, respectively.  

Details of our imaging setup are provided in Ref.\,\cite{sohmen2021birth}.  In-trap and TOF images are performed along the vertical direction (along $B$ and gravity), using standard phase-contrast and absorption techniques, respectively.  The resolution of our in-trap images is approximately one micron.  We use a $36$\,ms TOF duration for imaging interference patterns.  

\noindent \textbf{Atom number:} We extract the condensed atom number $N$ from absorption imaging performed along a horizontal direction in a separate set of experimental trials under otherwise identical experimental conditions.  This allows for a larger field of view, and better fitting of thermal atoms.  $N$ is determined by subtracting the fitted thermal component from the total absorption signal.  

For comparison between experiment and theory, and between the variational and eGPE theory methods, we associate $N$ with the number of atoms in the droplets, and not in the diffuse halo that surrounds the droplets.  From simulation of TOF expansion, we find that the halo is repelled at early expansion times, and is likely indistinguishable from the thermal cloud in our TOF measurements.  While it is possible that some of the halo is counted in $N$, we neglect this possibility and assume that $N$ includes only atoms within droplets.  

\noindent \textbf{Scattering length:} The positions of phase boundaries between different droplet configurations are quite sensitive to the scattering length $a$, which is not known with high precision in our range of magnetic fields.  For all theory, we use a value of $a = 88\,a_0$, where $a_0$ is the Bohr radius, as this value provides good agreement between experiment and theory for the 1D-to-2D transition point.

\noindent \textbf{Extracting critical aspect ratio:} The critical aspect ratio $\alpha^*_t$ is extracted from fit to the function $\alpha_a=\alpha_{0}$ for $\alpha_t<\alpha^*_t$, $\alpha_a=\sqrt{\alpha_{0}^2 + b(\alpha_t-\alpha^*_t)^2}$ for $\alpha_t>\alpha^*_t$, where $\alpha^*_t$, $\alpha_{0}$, and $b$ are fit parameters.  The error bars reported in Fig.\,\ref{fig:2}c represent the standard error on the mean, and are smaller than the markers on most points.  

\noindent\textbf{Interference patterns:}
The predicted interference patterns of Fig.\,\ref{fig:4} are calculated by assuming free expansion of Gaussian droplets.  In reality, the droplets are probably not Gaussian, and interactions during TOF expansion may modify the interference pattern.  However, the droplet shape primarily effects the envelope of the interference pattern, which is not our primary interest here, and from eGPE simulations, we expect the effects of interactions to be minor, provided that the droplets become unbound in a time short compared to the TOF, which we verify by both looking at shorter TOFs and comparing the fringe spacing observed in TOF with that expected from the in-trap droplet spacing.  The positions and size of the droplets are tuned to provide illustrative interference patterns.  

\noindent\textbf{Droplet number:}
We extract the droplet number from our in-trap images using a peak-finding algorithm applied to smoothed images.  The algorithm finds the local maxima above a threshold, which is chosen to be $40\%$ of the overall peak value. Each in-trap density distribution is classified as linear array or 2D zig-zag based on the atomic aspect ratio. Finally, the counts with a given droplet number are normalized by the total number of trials to get the probability shown in Fig.\,\ref{fig:2}d.  Fluctuations in the number of atoms in a given trial can push droplets above or below the threshold value, contributing to the spread in extracted droplet number for a given $\alpha_t$.  


\end{document}